# Broadband phase-preserved optical elevator

**Yuan Luo,[1,2,3]\* Tiancheng Han,[4] Baile Zhang,[2] Cheng-Wei Qiu,[4]\***

**and George Barbastathis[1,2]**

[1]*Department of Mechanical Engineering, Massachusetts Institute of Technology,*

*Cambridge, Massachusetts, USA, 02139*

[2]*Singapore-MIT Alliance for Research and Technology (SMART) Centre,*

*Block S16-06-17, 3 Science Drive 2, Singapore 117543*

[3]*Center for Optoelectronic Biomedicine, College of Medicine, National Taiwan University,*

*Taiwan R.O.C., 10051*

[4]*Department of Electrical and Computer Engineering, National University of Singapore,*

*Singapore, 119620*

\* Email: yuan_luo@ntu.edu.tw; eleqc@nus.edu.sg

Abstract

Phase-preserved optical elevator is an optical device to lift up an entire plane virtually without distortion in light path or phase. Using transformation optics, we have predicted and observed the realization of such a broadband phase-preserved optical elevator, made of a natural homogeneous birefringent crystal without resorting to absorptive and narrowband metamaterials involving time-consuming nano-fabrication. In our demonstration, the optical elevator is designed to lift a sheet upwards and the phase is verified to be preserved always. The camouflage capability is also demonstrated in the presence of adjacent objects of the same scale at will. The elevating device functions in different surrounding media over the wavelength range of 400-700 nm. Our work opens up prospects for studies of light trapping, solar energy, illusion optics, communication, and imaging.

Transformation optics and conformal mapping [1,2] have provided new methods and insights to better manipulate light propagation using invariance of Maxwell's equations [3,4]. Through transformation optics, a variety of conceptual devices have been designed, such as perfect lens [5], hyper lens [6-9], and invisibility cloaks [1,2,10-14]. Recently, electromagnetic devices in expansive manners [15-19] were proposed for applications such as solar concentrators [20,21], communication components [22], and novel imaging devices [16,23,24]. Unlike invisibility cloaking, rather than compressing an object into either a point or a sheet, these devices can extend the object's boundary beyond its original space. However, these devices proposed thus far rely on metamaterials [15-19,21,22] or plasmonic resonant structures [25,26], which are typically narrow-band and sub-wavelength operations. In addition, metamaterial realization in the optical regime requires expensive and time-consuming nano-fabrication process to accurately achieve sub-wavelength features due to extreme values of constitutive parameters of materials, and more critically they are absorptive [18,27,28]. Hence, practical realization of electromagnetic devices in expansive manners at optical frequencies has still remained elusive.

On the other hand, lifting a macroscopic object physically is not surprising at all. One can lift it by applying either simply mechanical force or using sculpted light fields [29], in an analog to aerodynamic lift. However, the lifting force from using sculpted light heavily depends on the asymmetry of the refractive objects, so as to create lifting force from momentum conservation. Even when the object is asymmetric and the upward force exists, it can be understood straightforwardly that the lifting effect strongly relies on the lightness of the object, to compete with the gravitational force. Moreover, such lifting effect is not stable, because there are forces along other directions in addition to the upward direction. Then along the lifting path, the object will rotate and follow an unstable trajectory. The object may be rotated to a certain position where the upward force no longer exists [29], and then downward force may pull it back again.

In this connection, we propose a distinct mechanism to showcase the lifting effect without applying any mechanical force or altering the object's physical position. More exactly, the object stands still, and the lifting effect is optically pronounced by using an optical elevating device, which is non-metamaterial based. It is important to note that the optical elevation does not alter the optical path or phase, so that the optical elevation is rigorously perfect toward external observers. Our design adopts affine expansive transformation to expand the electromagnetic space vertically such that the entire sheet can be elevated dynamically with preserved optical path and phase in broad frequency band and wide viewing angles. We report and experimentally demonstrate the first realization of such a phase-preserved optical elevator, lifting a sheet underneath upwards optically while the sheet is still at the original height. This concept also provides further ability to conceal exterior hardware in the surrounding environment, presenting great advantages in camouflage for industrial and military purposes.

In our demonstration, the phase-preserved optical elevator is made of a single piece of natural birefringent calcite crystal. Thus, there is no need for nano-fabrication. Moreover, our approach guarantees low-loss and broadband operation at optical frequencies within different surrounding media such as air and oil. The phase-preserved elevating effect was directly presented using laser beams in green, blue, and red colors. The arbitrary range of elevation can be dynamically achieved by simply stacking multiple phase-preserved elevating devices.

A schematic of the phase-preserved elevating device is shown in Figure 1. The phase-preserved elevator has a cuboidal shape (Fig.1a) to lift a sheet underneath (Fig. 1b) upwards with an elevated height $h$. At the same time, this device can create a camouflage that matches the surrounding environment. The geometrical design parameters are labeled in Figure 1a, and the refractive indexes of extraordinary ($n_e$) and ordinary ($n_o$) polarized lights are ~1.65 and 1.48. The top and bottom surfaces are polished with fine optical surface quality.

The device must preserve light path without distortion in both physical and transformed virtual space, as illustrated in respective Figure 2a and b. Figure 2a illustrates a ray-trace in transverse-magnetic (TM) field polarization as a test on the performance of the design. In the physical space, a sheet under the elevating device is mapped to a lifted virtual space with the increased height of $h$. In Figure 2, it can clearly be seen that on the observer plane with the same incoming light, owing to the presence of the phase-preserved elevating device, the image of reflected beam through the device are identical to the image of a beam reflected directly by an elevated space without the device.

To implement the design of the phase-preserved optical elevator, let us first consider the following transformation from the original coordinates ($x'$, $y'$, $z'$) into a new set of coordinates ($x$, $y$, $z$):

$$x = x'$$
$$y = \frac{H}{H-h}y' + \frac{H}{h-H}h \quad (1)$$
$$z = z'$$

Such a transformation expands the electromagnetic space. Therefore, through this transformation, we can obtain the following constitutive parameters for the material [5,6] as

$$\varepsilon = \mu = \frac{J \cdot J^T}{\det(J)} = \begin{pmatrix} H-h/H & 0 & 0 \\ 0 & H/H-h & 0 \\ 0 & 0 & H-h/H \end{pmatrix} \quad (2)$$

where $J$ is the Jacobian of the transformation between the virtual space ($x'$, $y'$, $z'$) and physical space ($x$, $y$, $z$).

In order to make the material realizable in optical frequencies, here we consider two-dimensional (2D) geometry with the vertical polarization, i.e., the magnetic field can only interact with the $z$–component of the permeability tensor. We then make this component to be unit such that a nonmagnetic phase-preserved elevator can be created. Accordingly, the other components of the permittivity tensor need to be adjusted to keep the refractive index unchanged. Therefore, we can adopt the following permittivity tensor in the 2D case.

$$\varepsilon = \begin{pmatrix} (H-h/H)^2 & 0 \\ 0 & 1 \end{pmatrix} = diag\left[\left(\frac{H-h}{H}\right)^2, 1\right] \quad (3)$$

This means that the permittivity can be expressed as two values in two principal directions, respectively. Consequently, we can obtain the principal values of the permittivity as about $1.48^2$ and $1.65^2$, if we choose the refractive index of background to be 1.65. This particular material can be simply realized using a single natural calcite crystal with $n_x = n_o = 1.65$ and $n_y = n_e = 1.48$.

We next experimentally demonstrate the performance of the phase-preserved elevating device in different media. This experiment is schematically illustrated in Figure 3a. To prove the elevating effect, we fill a glass tank with a transparent laser liquid (OZ-4Laser IQ, CODE 5610, from Cargille Labs, n ~ 1.55 measured at visible wavelength). To better visualize the effect of phase-preserved elevation, a mask pattern "S•G", printed on an opaque plastic plate with thickness of 1 mm using a stereolithography machine (ViperTM SLA System), is located in front of the laser-head. The position of the pattern is placed such that the TM polarized light transmitting via the symbol "S•" will go through the phase-preserved elevator with a mirror sheet underneath. Meanwhile the light that goes through the symbol "G" will be reflected by a mirror sheet's surface directly, which is placed at the height of $h$=2.3mm with volumetric target objects underneath. A CCD (charge-coupled-device) camera is used to monitor the projected image from the opposite side of the tank. If the phase-preserved elevator can camouflage the target objects successfully, the CCD camera will image a well-patterned "S•G" as though the mirror sheet under the phase-preserved elevator were lifted up to the same height ($h$) of the camouflaged target objects.

This behavior exactly was confirmed by measurements at three different incident wavelengths (488nm, 561nm, 650nm) and at various angles. Figure 3b shows the image obtained when the light through "S•" was reflected from the plain mirror sheet surface with the incident angle ($\theta$) of 65°. In the image, the symbols "S•" and "G" are located at different altitudes, in consistence with the different positions of two mirrors (as Fig. 3a shows) in the very beginning. In contrast, Figure 3c shows the case when "S•" is reflected from the phase-preserved elevated mirror surface, all letters in the captured CCD image are located at the same height, serving as the direct evidence of virtually elevating a lower surface into a higher mirror sheet at the desired height.

Each single device has limited elevation distance, which relies on the thickness of the natural birefrigent crystal. However, such birefringent crystal is usually thin in nature. In order to achieve arbitrary elevation, we simplify the realization of the phase-preserved elevating effect at will by stacking the phase-preserved elevators. Figure 3d shows the experimental image when there are no phase-preserved elevators along "S•" light path while the light through "G" is reflected from the mirror sheet surface with doubled height, $h$=4.6mm. Hence, the symbol "G" is located further away from "S•" (compared with Fig. 3b). Figure 3e shows the results when two phase-preserved elevators are stacked above the mirror sheet along the light path of "S•", and thus all letters in the captured CCD image are located again at the same altitude.

In addition, the phase-preserved elevating effect was confirmed not only in liquid medium but also in air (Fig. 3f and g). Instead of using liquid, the surrounding medium in the tank was replaced by air. Through the same procedure as described above, Figure 3f shows the image without the phase-preserved elevator along the light path through "S•" while the light through "G" is reflected from the mirror surface with increased height, $h$~9.3mm, resulting in the pattern "G" appearing outside the CCD region. Nevertheless, as seen in Figure 3g, when the phase-preserved elevator is put in the light path via "S•", all the patterns on the CCD plane locate at the same altitude, as expected.

To verify quantitatively that our phase-preserved optical elevator functions independently of incident angle, we provide further studies at three additional angles of 60°, 70°, and 80° shown in Figure 4. All letters in the captured CCD image are located at the same height. This serves as direct evidence of the phase-preserved elevator to preserve light path in the expanded space without distortion in broadband operation over a large range of various viewing angles.

To unambiguously prove the phase-preserved feature with broadband operation, we imaged interference patterns [30-32] with different wavelengths in air using a Mach-Zehnder interferometer with phase-preserved elevator in one arm and free-space as reference. A schematic of the measurement is presented in Figure 5a. The interference patterns of Figure 5b exhibit exactly the same number of fringe patterns in green, blue, and red colors, thus successfully demonstrating broadband operation in at least the visible regime. Subsequently, the phase-preserved elevating device is replaced by a mirror at an elevated height of 9.3mm. As shown in Figure 5c, the interference patterns are identical. Hence, the optical path in the lifted, expanded space has remained invariant upon introducing the phase-preserved elevator.

In summary, we have successfully demonstrated phase-preserved elevator at optical frequencies using a natural birefringent crystal. Our design scales to arbitrary elevation height by stacking sheets of calcite, and is invariant to color within the visible range. Moreover, the optical elevator can simultaneously be used to camouflage arbitrarily large exterior objects. The elevation height and camouflage volume are far beyond the capability of metamaterials. The proposed design with robust function and simple fabrication presented here may also further facilitate future three-dimensional expanding devices as well as other transformation based optical devices.

## Acknowledgements

We acknowledge financial support from the National Institutes of Health (RO1CA134424), the BioSystems and BioMechanics (BioSyM) Independent Research Group of the Singapore-MIT Alliance for Research and Technology (SMART) Centre (Contract No. 015824-039). Y.L. acknowledges a fellowship from Taiwan's National Science Council (Contract No. NSC-97-2917-I-564-115). C.W.Q. acknowledges the support from the National University of Singapore (Grant No. R-263-000-574-133). T.H. acknowledges a financial support from China Scholarship Council (CSC) for his joint PhD in National University of Singapore.

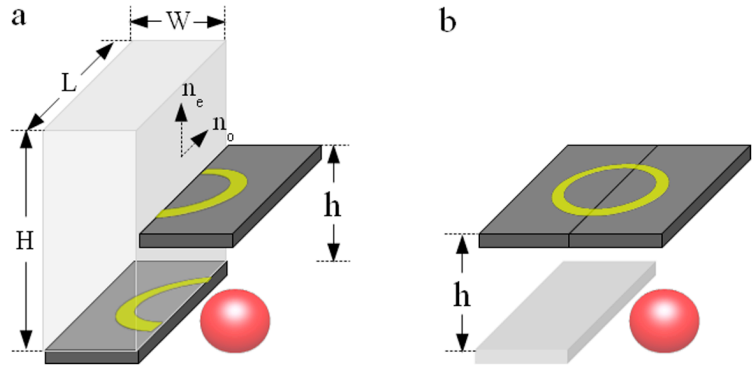

Figure 1: **(a,b)** A phase-preserved elevating design virtually floats a sheet, enlarges the volumetric space under the sheet, and further provides the camouflage capability of exterior environment. The cuboidal phase-preserved optical elevator is made of a single piece of natural birefringent crystal with the geometrical parameters of H=19.7mm, L=40mm, and W=10mm.

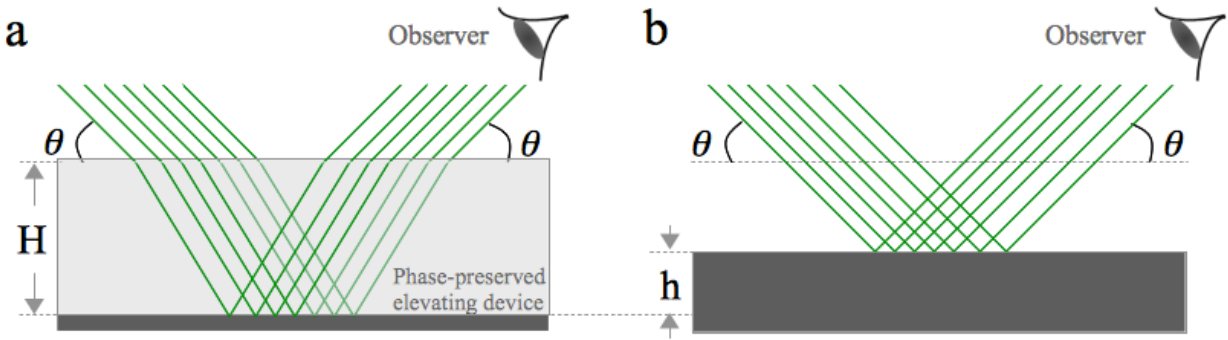

Figure 2: Illustrations of a phase-preserved optical elevator, performed with ray tracing. Light at 561nm with the incident angle of $\theta$ passing through the phase-preserved elevator system (**a**) is identical to its corresponding expansive virtual space without phase-preserved elevators (**b**). The observer in both **a** and **b** is located at the same position with respect to the identical angle.

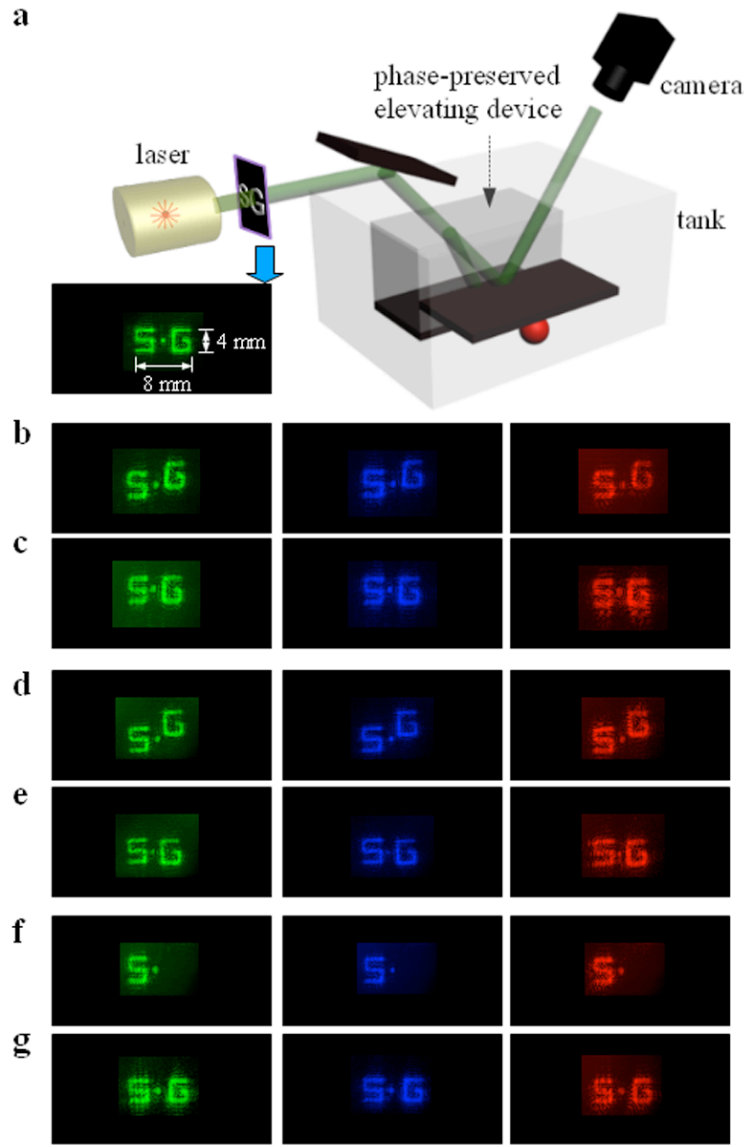

Figure 3: Phase-preserved optical elevating using green, blue, and red laser beams. **a,** Schematic diagram of the experimental setup. The laser beam goes through a mask pattern "S•G", and a CCD camera is monitoring the phase-preserved elevating effect above the tank. The surrounding medium in the glass tank is first filled with a transparent liquid of refractive index ~1.55 for measurement **b-e**, and then replaced by air for **f-g**. (**b,c**) The images captured on the CCD camera when the light through "G" is reflected by a mirror sheet placed at a height of $h$=2.3mm. Meanwhile, light through "S•" is reflected from the plain mirror sheet without the phase-

preserved elevator (b), and with the phase-preserved elevator above the plain mirror (c). (**d,e**) The resultant images shown on the CCD camera when the light through "G" is reflected by a mirror sheet at a doubled height of $h$=4.6mm. Meanwhile, light through "S•" is reflected from the plain mirror sheet without the phase-preserved elevator (d), and with the phase-preserved elevator above the plain mirror (e). (**f,g**) Light through "G" is reflected by a mirror sheet with height of $h$=9.3mm. Meanwhile, light through "S•" is reflected from the plain mirror sheet without the phase-preserved elevator (f), and with the phase-preserved elevator above the plain mirror (g).

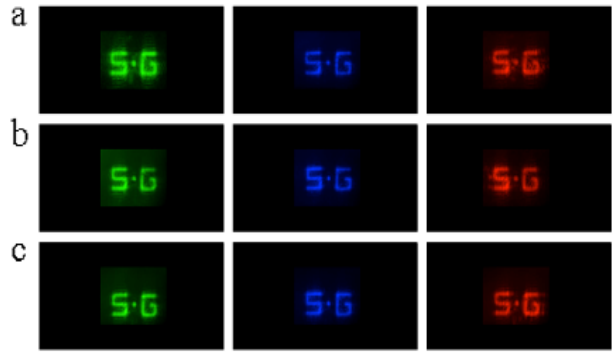

Figure 4: Images of "S•G" captured on the CCD with different incident angle of (**a**) 80°, (**b**) 70°, and (**c**) 60°.

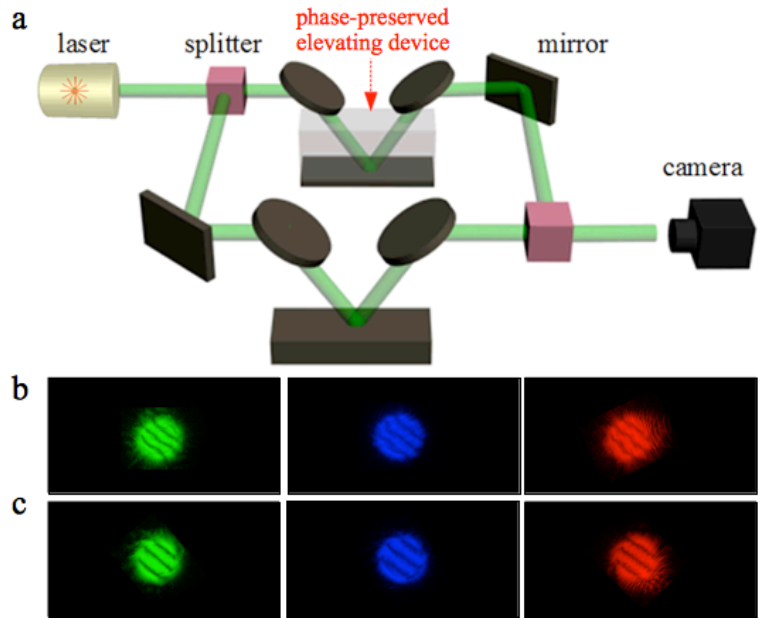

Figure 5: Optical characterization of the phase-preserved optical elevating in free space using Mach-Zehnder interferometer at wavelengths of 561nm, 488nm, and 650nm. **a,** Schematic of the experimental interferometer setup with the phase-preserved elevator in one arm and free-space as reference. **b,** The images captured on the CCD show exactly the same number of interference patterns for all three colors, unambiguously demonstrating broadband operation in the visible regime. **c,** The phase-preserved elevator is replaced by a mirror placed at an elevated height of 9.3mm. The images captured on the CCD are identical to the patterns in **b**, demonstrating the optical path and phase preserved without distortion in the corresponding expanded space.